\documentclass[]{aa} 
\usepackage{graphicx}
\begin{document} 

%\thesaurus{6(06.01.2, 06.18.1, 06.13.1, 06.18.2)} 
\title{
Periodicities in data observed during the minimum and the rising phase of solar cycle
23;  years 1996 - 1999.
}

\author{S. Zi\c{e}ba, J. Mas{\l }owski, A. Michalec and A. Ku\l ak\inst{}}
\institute{Astronomical Observatory, Jagiellonian  University,  ul  Orla 
171, PL30-244 Krak\'ow, Poland 
} 
\offprints{A. Michalec (michalec@oa.uj.edu.pl)} 
\mail{michalec@oa.uj.edu.pl}
\date{Received 10.07.2001/ accepted 31.07.2001} 

\titlerunning{Periodicities in solar data}
\authorrunning{S. Zi\c{e}ba et al. }

%\maketitle

\abstract{ 
Three types of observations: the daily values of the solar radio flux 
at 7 frequencies, the daily international sunspot number and the daily Stanford 
mean solar magnetic field were processed in order to find all the periodicities 
hidden in the data. Using a new approach to the radio data, two time series were 
obtained for each frequency examined, one more sensitive to spot magnetic fields, 
the other to large magnetic structures not connected with sunspots. Power 
spectrum analysis of the data was carried out separately for the minimum  (540 days 
from 1 March 1996 to 22 August 1997) and for the rising phase (708 days from 23 
August 1997 to 31 July 1999) of the solar cycle 23. The Scargle periodograms obtained, 
normalized for the effect of autocorrelation, show the majority of known periods 
and reveal a clear difference between the periodicities found in the minimum and 
the rising phase. We determined the rotation rate of the `active longitudes' 
in the rising phase as equal to 444.4 $\pm$ 4 nHz ($26\fd0 \pm 0\fd3$). 
The results indicate that appropriate and careful analysis of daily radio data at 
several frequencies allows the investigation of solar periodicities generated in 
different layers of the solar atmosphere by various phenomena related to the periodic 
emergence of diverse magnetic structures.
}

\maketitle 

\section{Introduction} 

The study of periodicities in different solar data is important for the understanding 
of solar magnetic activity. For over a decade many authors have reported various 
periods other than those at 11 yr and 27 days, of which the one near 154 days is the 
best known (Rieger et al. \cite{rie84}; Dennis \cite{den85}; Bogart \& Bai \cite{bog85};
 Lean \& Brueckner \cite{lea89}; Bai \& Cliver \cite{bai90}; Dr\"{o}ge et al. \cite{dro90}; 
Pap et al. \cite{pap90}; Carbonell \& Ballester \cite{car90}, \cite{car92}; 
Bai \& Sturrock \cite{bai91}, \cite{bai93}; Kile \& Cliver \cite{kil91}; 
Bouver \cite{bou92}; Oliver et al. \cite{oli98}; Ballester et al. \cite{bal99}). 
This investigation was designed to show that  accurate daily radio observations 
can be very useful for studying the solar periodicities generated in different layers 
of the solar atmosphere, and in different phases of the solar cycle. We used a new 
approach to the daily measured radio fluxes which allow us to separate to some extent 
the radio emission generated in the strong magnetic fields of active regions from that 
emitted by large but weaker magnetic structures. All the solar data from the minimum 
and the rising phase of solar cycle 23 were processed separately using the Scargle 
periodogram technique (Scargle \cite{sca82}). The statistical significance levels of all the 
periods found were estimated through the $FAP$ (false alarm probability -- the probability 
that the given periodogram value $z$ is generated by noise), as well as by a Monte Carlo 
approach to the periodograms obtained (Horne \& Baliunas \cite{hor86}; Bai \cite{bai92b}; 
\"{O}zg\"{u}\c{c} \& Ata\c{c} \cite{ozg94}; Oliver \& Ballester \cite{oli95}). 

Papers published by Das \& Chatterjee (\cite{das96}); Das \& Nag (\cite{das98}, 
\cite{das99}) presented some periods in radio data without any discussion about 
their significance level. Our analysis of radio observations however, 
gives the majority of the known periods, reveals a clear difference between 
periodicities observed in two phases of the solar cycle and shows that daily 
measured radio fluxes at various frequencies are very useful for the systematic study 
of solar periodicities observed in the different layers of the solar atmosphere.
	  
\begin{table*}[t]
\caption{Parameters of the two different models of the solar radio emission used to eliminate 
the basic component $B$ from the daily observed flux $F_i$.}
\label{tab-param}
\begin{flushleft}
\begin{scriptsize}
\begin{tabular}{lccccccc}

\hline\noalign{\smallskip}
\\
\multicolumn{8}{l}{Linear model                    $F_i = B + h(ISN)_i$}\\
\\
\hline
frequency in MHz    &      405    &  810  &  1215  &  1620  &   2800     &       4995    &        8800\\
\hline
$B$- basic component  [su]  &  22.9 $\pm$.2  &  39.8 $\pm$.3  & 48.2 $\pm$.4  &  53.5 $\pm$.4  & 67.3 $\pm$.5  &  117.0 $\pm$.5  &    199.4 $\pm$.1 \\
$h$- radio flux production   &        0.138  $\pm$.003   &        0.337   $\pm$.005     &      0.508   $\pm$.007   &        0.611   $\pm$.007     &      0.785   $\pm$.008     &       0.633    $\pm$.008  &       0.752   $\pm$.017\\
\%  of  the data variance   &  71.0    &       78.6  &  81.9  &  85.7   &  87.5    &     83.5  &    61.6\\
explained by the model\\
\hline
\\
\multicolumn{8}{l}{Boltzman formula model  ~ $F_i = B_i + h_\mathrm B (ISN)_i = A_2 +(A_1 - A_2 ) / (1 + \exp ((t_i - t_\circ) / \Delta t ))+h_\mathrm B (ISN)_i$}\\
\\

\hline
frequency in  MHz      &    405   &   810  &  1215   & 1620   &  2800    &        4995   &         8800\\
\hline
parameters of~~    $A_1$   [su] &   22.1 $\pm$.5  &       38.9 $\pm$.8  &      48.0 $\pm$1   &     53.1 $\pm$1.2   &      67.4 $\pm$ 2  &    117.8 $\pm$ 1.4   &  200.5 $\pm$ 7.6\\
the Boltzman~~   $A_2$  [su]      &    31.4 $\pm$.5    &     62.5 $\pm$.7   &     75.0 $\pm$.9   &    78.9 $\pm$ 1.1   &   100.7 $\pm$ 1.8    & 137.1 $\pm$ 1.5  &   292.5 $\pm$ 7.1\\
sigmoidal~~~~~   $ t_{\circ}$ [days]   & 807 $\pm$15   &       855 $\pm$ 7  &   879 $\pm$ 8  &  880 $\pm$ 10  & 926 $\pm$ 12  &  962 $\pm$ 11  &  1064 $\pm$ 20\\
formula~~~~~~  $\Delta t$ [days] & 132 $\pm$ 13   &       111 $\pm$ 7  &         80 $\pm$ 7   &      91 $\pm$  9   &     106 $\pm$ 12    &     45 $\pm$ 12     &    143 $\pm$ 16\\
$h_\mathrm{B}$- radio flux production    &     0.067  $\pm$.004  &   0.154   $\pm$.006    &       0.291   $\pm$.008     &      0.408   $\pm$.010   &         0.542    $\pm$.012     &     0.487  $\pm$.012    &      0.261    $\pm$.025\\
\% of the data variance  &  81.0   & 91.8  &  91.4  &  91.5   &  92.5   &          87.5   &        79.3\\
explained by the model\\

\noalign{\smallskip}
\end{tabular}
\end{scriptsize}
\end{flushleft}
\end{table*}

\section{Observations }
The radio fluxes at the four lowest frequencies  405, 810, 1215, 1620 MHz 
are from our new accurate observations, 
which we began on 1 March 1996 (http://www.oa.uj.edu.pl.). 
Every day we measure the total solar radio flux at ten frequencies 
within the decimeter range of wavelengths using the 8-m radiotelescope built 
in 1995 at the Cracow Astronomical Observatory (Zi\c{e}ba et al. \cite{zie96}).
The international sunspot numbers $(ISN)$, the Stanford mean solar 
magnetic field $(MMF)$ and radio fluxes at the frequencies 2800, 4995
 and 8800 MHz were taken from Solar Geophysical Data bulletins (\cite{sol96}). 
We added to our radio observations those at higher frequencies for two reasons. 
Firstly, we wanted to examine periodicities of the radio emission in a wide enough 
radio band to analyse the radio flux coming from a large part of the solar 
atmosphere and secondly to compare our observations with those of others. 

It is trivial to form the time series from the daily values of the sunspot 
numbers and the mean magnetic field, but if we want to use the radio data for the 
study of magnetic activity, we must first eliminate from the observed daily flux 
(free from bursts) the thermal emission which comprises the majority of the daily 
measured value. This can be done through a procedure in which the flux observed 
on day $t_i$, $F_i$ is divided between the thermal, almost constant component (often 
called the basic component - $B$) and the slowly varying component $(SVC)$, whose
value changes every day and is generated by mechanisms dependent on the magnetic 
field. It is usually assumed that this component is proportional to a certain daily 
index of activity, for example $(ISN)_i$ and then the daily flux $F_i$  can be described 
by the following linear formula:  $Fi = B + (SVC)_i  =  B + h (ISN)_i$ , where 
$h$ is the production of the radio flux from the spot with  $ISN = 1$ 
(Kr\"{u}ger \cite{kru79}). 

The assumption that $B$ is constant over the large time intervals is rather strong 
and cannot be accepted, especially at a time when the level of radio flux rises 
systematically. Here we propose a new approach in which the basic component 
$B$ is not constant over the given time interval but changes every day according to 
the Boltzman sigmoidal formula:

       $$ B_i = A_2 + (A_1 - A_2) / (1 + \exp ( (t_i -t_{\circ}) / \Delta t) )$$ 
where $A_1$,  $A_2$, $t_{\circ}$, $\Delta t$  are the parameters and $t_i$ is the day number. 

\begin{figure}
\resizebox{\hsize}{!}{\includegraphics{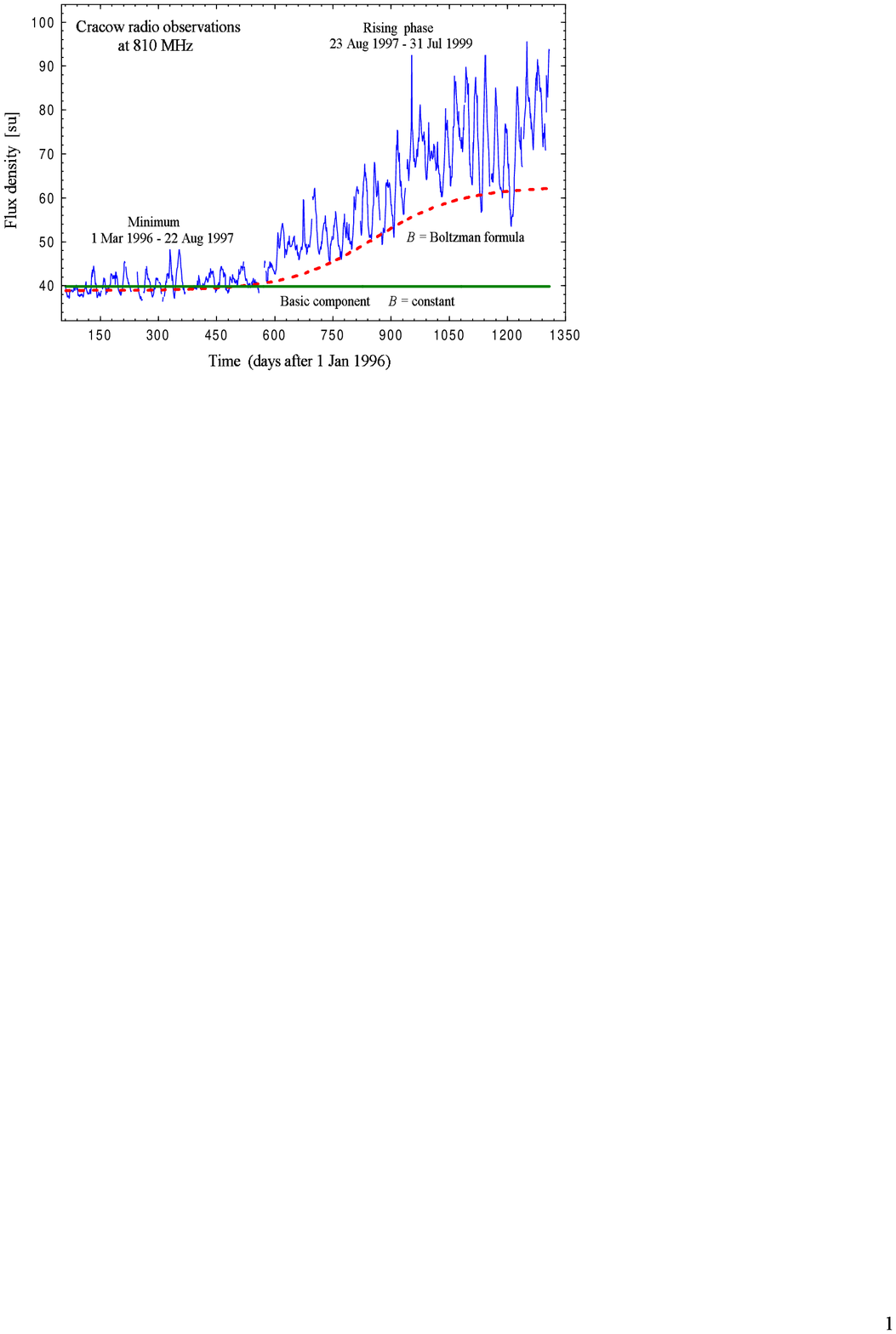}}
\caption{
Daily values of the radio flux at 810 MHz observed from Cracow. The horizontal solid 
line shows the constant value of the basic component  $B = 39.8$ su  resulting from the 
linear formula, while the dashed curve shows values of the basic component calculated 
from the best fitted parameters,  $A_1=38.9$ su, $A_2=62.5$ su, $ t_{\circ}=855$ days,
 $ \Delta t=111$ days according to the Boltzman formula 
 $B_i=A_2+(A_1-A_2) / (1+\exp ((t_i-t_{\circ}) / (\Delta t))$.
The division into the minimum and rising phase is also indicated.
}
\label{per1}
\end{figure}

\begin{figure*}
%\vspace{12cm}
\resizebox{\hsize}{!}{\includegraphics{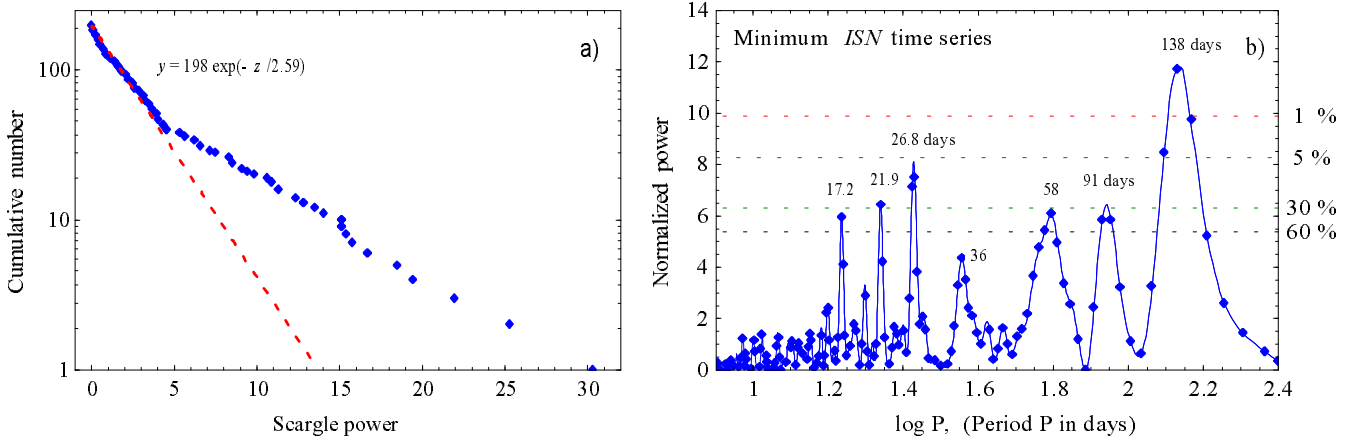}}
\caption{
a). The cumulative distribution function of the Scargle power for the original, 
minimum $ISN$ time series. The vertical axis is the number of frequencies whose power 
exceeds $z$. The straight line is the best fit to the points for values of power lower 
than 5. b). The normalized periodogram of the original, minimum $ISN$ time series with 
$FAP$ significance levels indicated.
}
\label{per2}
\end{figure*}

Then the observed daily flux   $F_i  = B_i + h_\mathrm{B} (ISN)_i$, 
and $h_\mathrm{B}$ has a similar interpretation to $h$. 
To determine the above parameters we used the observed radio data and daily 
sunspot  numbers  over  the whole time interval investigated,  1 March 1996 -
31 July 1999 (1248 days).The best fit values of these parameters are shown in 
Table 1. The difference between the two models is clearly seen, 
especially for four frequencies 405, 810, 1215, 8800 MHz. To demonstrate 
this we present in Fig. 1, as an example, the daily values of the radio flux 
at 810 MHz observed from Cracow as well as the calculated values of the basic 
component $B$ and $B_i$. The data in this figure also explain our division into 
 the minimum and the rising phase. 

Thus, in our approach to the radio data we create two time series from the 
observation at each frequency. The first, the $SVC$ (slowly varying component) time 
series, consists of diurnal values calculated as the difference between the daily 
observed flux and the daily value of the basic component computed from our model, 
$(SVC)_i = F_i - B_i$. The second, the $RRE$ (radio residual emission) time 
series describes the every day difference between the radio observations 
and our model of the daily radio flux, 
 $(RRE)_i = F_i - B_i - h_\mathrm{B} (ISN)_i$. Taking the time series 
$SVC$ and $RRE$, we can analyse cyclic variations of those magnetic structures 
which modified the observed radio emission. However, the $SVC$ series are more 
sensitive to spot magnetic fields, while the $RRE$ series are 
sensitive to large magnetic structures not connected with sunspots.
	
\begin{figure*}
%\vspace{6cm}
\resizebox{\hsize}{!}{\includegraphics{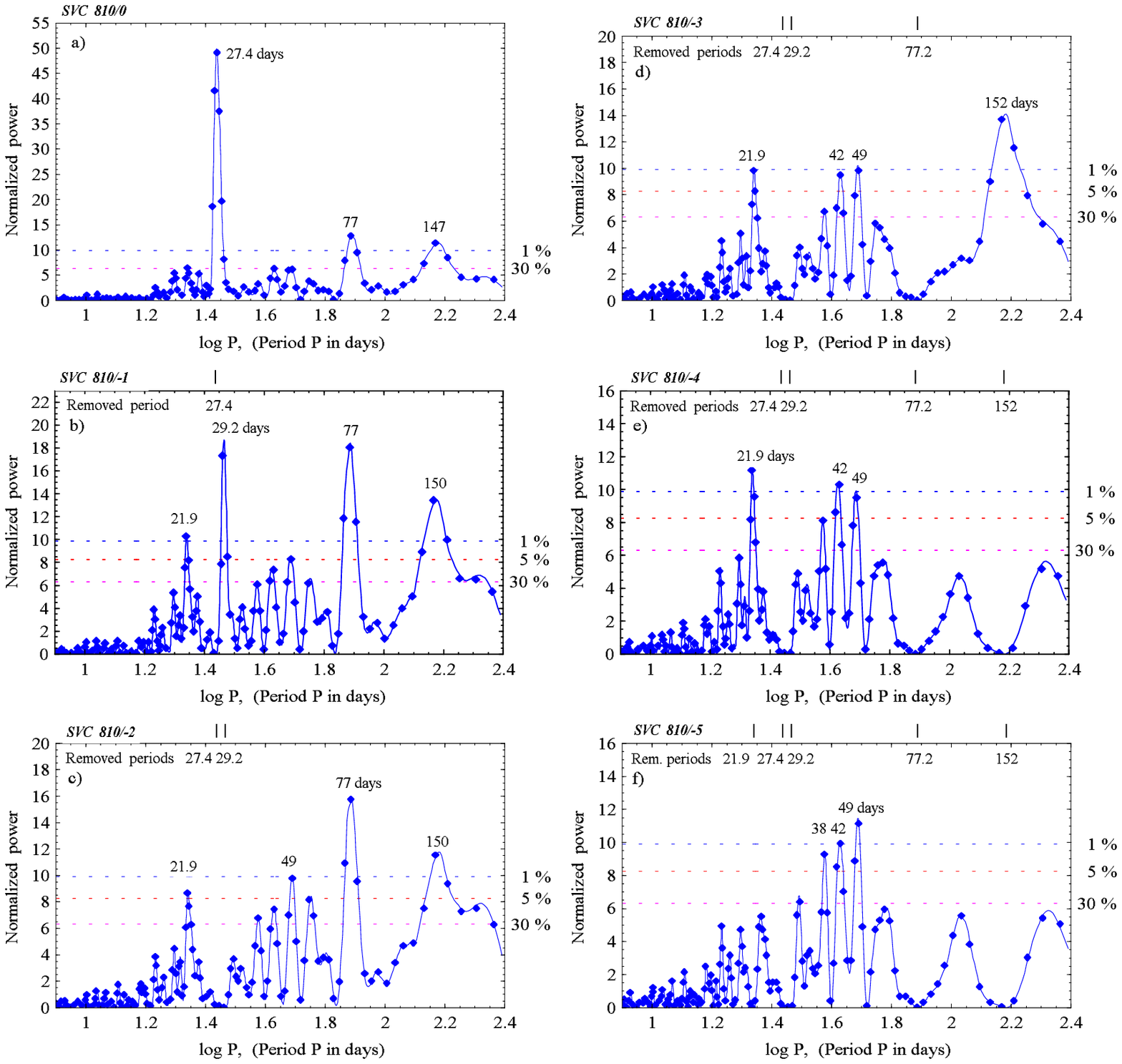}}
\caption{
a). The normalised periodogram of  the minimum, original $SVC~810/0$ time series. 
b), c), ... same as a) but recalculated after successively removing from the original data one, 
 two, and more sine curves having periods with peaks whose FAP values are smaller than 0.5\,\%. 
In each graph the removed periods are indicated at the top. The dashed lines show 
$FAP$ significance levels.
}
\label{det}
\end{figure*}
	
\section{Periodogram analysis}

The search for periodicities in all the time series was performed by calculating 
the Scargle normalized periodograms $P_N (\omega)$ (Scargle \cite{sca82}).This 
technique (see Horne \& Baliunas \cite{hor86}) has several advantages over the conventional fast 
Fourier transformation method and provides, through $FAP$, a simple estimate of 
the significance of the height of a peak in the power spectrum. However, the $FAP$ 
value is easy to calculate only for that time series for which the successive data 
are independent. In our case, the all analysed time series were prepared from 
the daily values of the different solar indices which are not independent but 
correlated with a characteristic correlation time of a week (Oliver \& Ballester \cite{oli95}). 
Therefore, for all our $FAP$ calculations used to estimate the statistical 
significance of a peak $z$, in the Scargle power spectrum, we applied  the formula 
   $FAP = 1- [1- \exp( - z/k )]^N   = 1- [1- \exp( - z_m)]^N$, where $N$ 
is the number of independent frequencies, $k$ is the normalization factor 
due to data correlation and $z_m = z/k$ the normalized power 
(Bai \& Cliver \cite{bai90}; Bai \cite{bai92b}). 
To determine the normalization factor $k$  for the given time series we 
followed the procedure described by Bai \& Cliver (\cite{bai90}). The key step of this 
procedure is the choice of a spectral window as well as the number of 
independent frequencies. Since we would like to analyse naturally limited time series 
connected with two different phases of the solar cycle over the largest possible 
range of periods, we took  the interval 43 to 1447 nHz (8-270 days)
as the spectral window for all the investigated time series. The shortest period, 8 days,
 is connected with a possible range of correlations of the analysed data 
(see Table 2 and 3 where the autocorrelation coeficients with the lag = 1 day 
and lag = 7 days are given).  The longest periods, 270 days,  
results from the actual length of the minimum time series equal to 540  days. 
The number of totally independent frequencies inside the chosen window is 
given by the value of the independent Fourier spacing, 
 $\Delta f_\mathrm{ifs} = 1/T$, where $T$ is the time span of the data 
(Scargle \cite{sca82}). In the case of our time series, $T_\mathrm{M} = 540$ days 
and $\Delta f_\mathrm{ifsM} = 21.433$ nHz  for the minimum, 
while for the rising phase $T_\mathrm{R} = 708$ 
days and $\Delta f_\mathrm{ifsR} = 16.348$ nHz. However, 
de Jager (\cite{dej87}) has shown by Monte-Carlo 
simulations that the Fourier powers taken at intervals of one-third of the independent 
Fourier spacing are still statistically independent. Thus, we accepted the numbers 
198 and 259 as the numbers of  independent frequencies in the chosen window 
for the minimum and the rising  phase time series, respectively.
 To illustrate the method used for determination of the normalization 
factor $k$, we will process one of our time series step by step; as an example 
we take the $ISN$ time series in the minimum phase. First, we calculate 
the Scargle power normalized by the variance of the data, 
(Horne \& Baliunas \cite{hor86}) for 
all 198 independent frequencies which allows us to construct the graph presented 
in Figure 2a. This shows the cumulative number of frequencies for 
which the Scargle power exceeds a certain value $z$. 
Then we fit  all values of power $z < 5$ to the equation  $y = 198 \exp(-z/k)$,  
which gives the value of the normalization factor $k$ equal 2.59, since when 
$\exp(-z/k) << 1$  the formula for $FAP$ reduces to  $FAP = N \exp(-z/k)$. Finally, we 
normalize the power spectrum once more by dividing the Scargle power by 2.59 to 
obtain the normalized periodogram for which $FAP$ values are easily calculated. 
Therefore, if we substitute $z = 30.54$, $k =2.59$, and $N = 198$ in the $FAP$ 
formula we get the normalized power $NP = z_\mathrm{m} = 11.79$ and 
$FAP = 0.0015$ for the highest peak in the $ISN$ minimum time series 
periodogram. Figure 2b presents this normalized 
periodogram of the minmum $ISN$ series together with $FAP$ significance 
levels obtained from $FAP$ formula with $N = 198$ and $k = 2.59$. 

\begin{table*}
\caption
{Charactaristic parameters for the minimum original time series(0) 
and all the next obtained after removing from the original data successively one,
two and more sine curves (`-1', `-2', ...) having periods with peaks whose $FAP$ values are smaller than 0.5\,\% These frequencies 
are printed in bold letters}
\label{tab2}
\begin{flushleft}
\begin{scriptsize}
\begin{tabular}{rrcrrrrrcrrcrrrrrc}
\hline\noalign{\smallskip}

  &\multicolumn{2}{c}{The  highest}   &\multicolumn{2}{c}{Auto}          &$k$ & $NP$ &$ FAP $ & \%  &   &\multicolumn{2}{c}{The  highest}   &\multicolumn{2}{c}{Auto}          &$k$ & $NP$ &$ FAP $ & \%  \\
     & \multicolumn{2}{c}{peak  at}    &\multicolumn{2}{c}{correlation}     &    &    & [\%]     & of  &      & \multicolumn{2}{c}{peak  at}    &\multicolumn{2}{c}{correlation}     &   &        &  [\%]      & of  \\
      & Freq.           &  Per.           &\multicolumn{2}{c}{lag}          &   &      &        &  var.  &       & Freq.           &  Per.           &\multicolumn{2}{c}{lag}          &   &      &        & var.  \\
      & [nHz]           &  [day]          &    1~   &   7~          &        &      &       &  exp.     &       & [nHz]           &  [day]          &    1~   &   7~                    &        &      &        &   exp.    \\
\hline
  1   &\multicolumn{2}{c}{2}              &\multicolumn{2}{c}{3}            &  4~     &  5~   &   6~    &    7~~  &   1   &\multicolumn{2}{c}{2}              &\multicolumn{2}{c}{3}            &  4~     &  5~   &   6~    &    7~~  \\
\hline
\multicolumn{2}{l}{\it ISN}&        &      &             &        &      &        &       &    \multicolumn{2}{l}{\it MMF}\\
$     ~0$ & {\bf 84a} & 138 & .839 & .126 & 2.59 & 11.8 & 0.15 & 17.4 &$ ~0$ & {\bf 858a} & 13.5 & .523 & -.187 & 1.56 & 19.0 & 0.00 & 12.6\\
             &             &      &      &      &      &      &       &      &$      -1$ & {\bf 419a}  & 27.6 & .476 & -.079 & 1.54 & 19.0 &  0.00 & 23.8\\
          &             &        &      &      &       &     &       &      &$      -2$ & {\bf 400a} &  28.9 & .390 & -.095 & 1.65 & 13.7 &  0.02 & 30.0\\
\multicolumn{2}{l}{{\it SVC~~405}} &     &     &    &    &      &        &       &\multicolumn{2}{l}{\it RRE~~405}\\
$     ~0$ & {\bf 418a} &  27.7 & .801 & .007 & 2.50 & 19.8 & 0.00 &  21.0 &$      ~0$ &  {\bf 416a} &  27.8 &  .714 &  .014 &  2.25 & 23.4  &  0.00  & 22.6\\  
\multicolumn{2}{l}{{\it SVC~~810}}         &    &     &    &    &      &        &       &\multicolumn{2}{l}{\it RRE~~810}\\
$     ~0$ & {\bf  423a} & 27.4  &  .905 & .177 & 1.56  & 49.9 &  0.00 & 35.9 &$       ~0$ & {\bf 418a} & 27.7 &  .801 &  .218 & 1.46 & 42.9 & 0.00 &  26.4\\
$     -1$ & {\bf 396a} & 29.2 &  .895 &  .285 & 1.71 & 18.7 & 0.00 &  45.2 &$       -1$ & {\bf 152a} & 76.2 & .765 &  .346 &  1.70 & 27.7  & 0.00 &  40.6\\
$     -2$ & {\bf 150b} & 77.2 &  .883 &  .321 & 1.89 & 15.6  &  0.00 &  52.1 &$       -2$ & {\bf 381a} & 30.4 & .707 &  .216 &  2.06 & 15.3 & 0.00 & 48.8\\
$     -3$ & {\bf 76b} & 152  &  .874 & .228 &  2.01 & 14.1 &  0.01 & 57.7 &$     -3$ & {\bf 268b} &  43.2 & .667 &  .242  & 2.07 & 11.8 &  0.14 &  54.2\\
$     -4$ & {\bf  529b}& 21.9 &  .852 & .105 &  2.10 & 11.2 &  0.28 & 61.9 &          &       &       &        &        &      &      &       &      \\
$     -5$ & {\bf  238b}& 48.6 &  .836 & .164 &  2.11 & 11.5 &  0.20 & 66.0 &          &       &       &        &        &      &      &       &      \\
\multicolumn{2}{l}{{\it SVC~ 1215}}          &    &     &    &    &      &        &       &\multicolumn{2}{l}{\it RRE~ 1215}\\
$     ~0$ & {\bf 424a} & 27.3 &  .912 &  .219 & 1.60 & 34.6 & 0.00 & 26.7 &$      ~0$ &  {\bf 157a} & 73.7 &  .757 &  .271 &  1.72 & 18.8 & 0.00 &  13.7\\
$     -1$ & {\bf 80b} & 145 &  .905 &  .297 &  1.72 & 14.8  & 0.01 & 34.5 &$       -1$ & {\bf 420a} &  27.6 &  .718 &  .160  & 1.91 & 18.7 &  0.00 &  26.7\\
$     -2$ & {\bf 151b} &  76.6 &  .893 &  .207  & 1.79 &15.0 &  0.01  & 41.8 &$     -2$ & {\bf  383b} & 30.2 &  .683 &  .221 &  1.93 & 12.2  & 0.10 & 34.3\\
$     -3$ & {\bf 528b} &  21.9 &  .886 &  .112  & 1.90 & 12.1  & 0.11  & 47.5 &$    -3$ & {\bf 267b}  & 43.4 &  .649 &  .233  & 1.89 & 11.8  & 0.14  & 40.6\\
$     -4$ & {\bf  113b}&  102 &  .873 & .176 &  1.94 & 11.5 &  0.20 & 52.9 &          &       &       &        &        &      &      &       &      \\
$     -5$ & {\bf  498b}& 23.2 &  .855 & .084 &  1.95 & 11.8 &  0.15 & 57.8 &          &       &       &        &        &      &      &       &      \\
$     -6$ & {\bf  242b}& 47.8 &  .839 & .129 &  2.18 & 11.8 &  0.15 & 62.5 &          &       &       &        &        &      &      &       &      \\
\multicolumn{2}{l}{{\it SVC~ 1620}}   &    &     &    &    &      &        &       &\multicolumn{2}{l}{\it RRE~ 1620}\\
$   ~0$ & {\bf 425a} & 27.2 &  .898 &  .111 &  2.15 & 23.1 &  0.00 & 25.0 &$        ~0$ & {\bf 155b} &  74.7 &  .687 &  .197 &  2.00 & 11.7 &  0.16 & 10.6\\
       &           &    &       &        &      &      &       &       & $       -1$ & {\bf 420b} &  27.6 &  .653 &  .123 &  2.07 & 11.2 &  0.28  & 19.3\\
       &             &      &       &       &        &     &       &       &$      -2$ &  {\bf 382b} & 30.3 &  .627 &  .159 &  2.12 & 10.8 &   0.40  & 27.2\\
\multicolumn{2}{l}{{\it SVC~ 2800}}      &    &     &    &    &      &        &       &\multicolumn{2}{l}{\it RRE~ 2800}\\
$     ~0$ & {\bf 85b} & 136 &  .931 &  .134 &  2.38 & 13.0 & 0.05 &  17.4 &$       ~0$ &  383b &  30.2 &  .647 &   .044 &  2.32 &  7.8 & 7.78  & 7.6\\
$     -1$ & {\bf 424a} & 27.3 &  .921 &  .020 &  2.57 & 13.5 & 0.03 & 28.0 &            &            &       &        &       &       &      &       &      \\
\multicolumn{2}{l}{{\it SVC~ 4995}}     &    &     &    &    &      &        &       &\multicolumn{2}{l}{\it RRE~ 4995}\\
$      0$ & {\bf 90a} & 129 &  .862 &  .286 & 2.00 & 18.3 &  0.00 & 14.3 &$       ~0$ & {\bf 53a} & 218 &  .710 &  .347 &  1.64 &23.5 & 0.00 &  19.4\\
$      -1$ & {\bf 124a}  & 93.3 &  .832 &  .166  & 2.20 & 13.9 & 0.02 & 24.6 &$       -1$ & {\bf 158a} & 73.2 & .625 &  .224 &  1.83 & 19.5 & 0.00 & 30.3\\
 \multicolumn{2}{l}{{\it SVC~ 8800}}  &    &     &    &    &      &        &       &\multicolumn{2}{l}{\it RRE~ 8800}\\
$      ~0$ & {\bf 122b} &  94.9 &  .635 &  .355 &  1.63 & 15.8 &  0.00 &  10.3 &$        ~0$ & {\bf 156b} &  74.2 &  .638 &  .396 &  1.52 & 16.2 &  0.00  & 10.0\\
$      -1$ & {\bf 154b} &  75.2 &  .593 &  .289  & 1.71 & 14.9 &  0.01 &  19.9 &$       -1$ & {\bf 121b} &  95.6 &  .598 &  .339 &  1.54 & 14.3  & 0.01  & 18.1\\
         &           &      &      &       &       &      &      &       &$       -2$ &  {\bf 83b}& 139 &  .558 &  .285 &  1.55 & 11.8 & 0.15 &  24.3\\

\noalign{\smallskip}
\hline
\end{tabular}
\end{scriptsize}
\end{flushleft}
The columns of the table show: 1: time series(`-1' indicates that the one sine curve with period given a row above was removed from the original data, `-2' indicates that the two sine curves with periods given in two rows above were removed from the original data, and so on), 2: frequency and period of 
the highest peak in a given time series, a small letter following a frequency value indicates to which 
interval of probability (`a': $<0.1$\,\%, `b': 0.1 - 1\,\%, `c': 1 -5\,\% ) this period belongs to
after the randomising procedure, 3: the autocorrelation coefficients calculated 
with two lags equal 1 and 7 days respectively, 4: the normalization factor $k$ computed 
according to the procedure described in Sect.\,3, 5: the normalized power equal to 
the Scargle power divided by $k$, 6: the $FAP$ value resulting from the normalized power, 
7: in successive rows a percentage of the original data variance explained by the 
prominent sinusoidal signals found in a given type of data.\\
\end{table*}

\begin{table*}
\caption{Same as Table 2 but for the rising phase data.}

\label{tab3}
\begin{flushleft}
\begin{scriptsize}
\begin{tabular}{rrcrrrrrcrrcrrrrrc}
\hline\noalign{\smallskip}
\hline

  &\multicolumn{2}{c}{The  highest}   &\multicolumn{2}{c}{Auto}          &$k$ & $NP$ &$ FAP $ & \%  &   &\multicolumn{2}{c}{The  highest}   &\multicolumn{2}{c}{Auto}          &$k$ & $NP$ &$ FAP $ & \%  \\
    & \multicolumn{2}{c}{peak  at}    &\multicolumn{2}{c}{correlation}     &   &    & [\%]     & of  &      & \multicolumn{2}{c}{peak  at}    &\multicolumn{2}{c}{correlation}     &   &        &  [\%]      & of  \\
      & Freq.           &  Per.           &\multicolumn{2}{c}{lag}          &    &      &        &  var..  &       & Freq.           &  Per.           &\multicolumn{2}{c}{lag}          &    &      &        & var.  \\
      & [nHz]           &  [day]          &    1~   &   7~          &        &      &       &  exp.     &       & [nHz]           &  [day]          &    1~   &   7~                    &        &      &        &   exp.    \\
\hline
  1   &\multicolumn{2}{c}{2}              &\multicolumn{2}{c}{3}            &  4~     &  5~   &   6~    &    7~~  &   1   &\multicolumn{2}{c}{2}              &\multicolumn{2}{c}{3}            &  4~     &  5~   &   6~    &    7~~  \\
\hline
\multicolumn{2}{l}{{\it ISN}}   &      &       &        &       &      &        &       &\multicolumn{2}{l}{\it MMF}\\
$~~0$ & {\bf 442a} &  26.2 &  .906 &  .150  & 2.19 & 19.2  &  0.00 &  44.6 & $ ~~0$ & {\bf 799a} &  14.5 &  .650 & -.415  &   1.54  &25.8  &    0.00 &  15.5\\
$      -1$ & {\bf 127a} & 91.1 &  .897 &  .179 &  2.37 & 13.9 &  0.02  & 49.7 &$  -1$ & {\bf 844a} &  13.7 &  .634 & -.340 &  1.70 & 26.9 &  0.00  & 30.1\\
        &             &       &       &        &      &       &      &       &$     -2$ & {\bf 876a} &  13.2  & .548 & -.187 &  1.93 & 12.2  &  0.14 &  36.4\\
\multicolumn{2}{l}{{\it SVC~~405}}   &    &     &    &    &      &        &       &\multicolumn{2}{l}{\it RRE~~405}\\
$~~0$ & {\bf 418a} &  27.7 &  .731 &  .086 &  1.90 & 24.5 &  0.00 & 21.9 &$ ~~0$ &  {\bf 418a} & 27.7 & .641 &  .064 &  2.15 & 15.2 &  0.01 &  9.8\\
$ -1$ & {\bf 55a} & 210 &  .695 &  .102 &  2.10 & 14.3 &  0.02 &  28.8 &$-1$ &  {\bf 52a} & 223 &  .609 &  .072  & 2.23 & 13.2  &  0.05 & 17.9\\
$    -2$ & {\bf 444a} &  26.1 &  .663 &  .017 &  2.15 & 11.1  &  0.39 &  33.9  &        &           &     &       &        &        &     &       &      \\
\multicolumn{2}{l}{{\it SVC~~810}}   &    &     &    &    &      &        &       &\multicolumn{2}{l}{\it RRE~~810}\\
$~~0$ & {\bf 443a} &  26.1 &  .935 &  .215 &  1.01 & 74.0 &  0.00 &  33.0 &$ ~0$ & {\bf 417a} &  27.8 &  .863 &  .206  & 1.30 & 52.0 &  0.00  & 20.2\\
$ -1$ & {\bf 416a} & 27.8 &  .924 &  .300 &  1.21 & 50.4 &  0.00 &  45.4 & $ -1$ & {\bf 450a} &  25.7 &  .842 &  .275  & 1.58 & 27.2 & 0.00 & 30.2\\
$ -2$ & {\bf 75a} & 154  & .918 &  .385  & 1.32 & 38.2  & 0.00 & 53.5 &$   -2$ &{\bf 56a} & 207 &  .824 &  .323 &  1.74 & 20.3 & 0.00 &  37.4\\
$ -3$ & {\bf 456a} & 25.4 &  .904 &  .275 &  1.51 & 20.1 &  0.00 &  58.0 &$    -3$ & {\bf 71b} & 163 &  .806 &  .245  & 1.91 & 13.2 & 0.05  & 42.0\\
$     -4$ & {\bf 62a  }& 187  &  .893 & .318 &  1.59 & 21.5 &  0.00 & 62.3 &          &       &       &        &        &      &      &       &      \\
$     -5$ & {\bf  103a}& 112 &  .878 & .239 &  1.67 & 21.4 &  0.00 & 66.4 &          &       &       &        &        &      &      &       &      \\
$     -6$ & {\bf  477a}& 24.3 &  .861 & .147 &  1.80 & 14.2 &  0.02 & 69.0 &          &       &       &        &        &      &      &       &      \\
$     -7$ & {\bf  385a}& 30.1 &  .853 & .193 &  1.74 & 15.4 &  0.01 & 71.5 &          &       &       &        &        &      &      &       &      \\
$     -8$ & {\bf  127b}& 91.1 &  .843 & .205 &  1.76 & 15.2 &  0.01 & 73.9 &          &       &       &        &        &      &      &       &      \\
$     -9$ & {\bf  562b}& 20.6 &  .829 & .143 &  1.77 & 12.8 &  0.07 & 75.6 &          &       &       &        &        &      &      &       &      \\
$     -10$ & {\bf  498b}& 23.2 &  .819 & .184 &  1.84 & 12.3 &  0.12 & 77.3 &          &       &       &        &        &      &      &       &      \\
\multicolumn{2}{l}{{\it SVC~ 1215}}   &    &     &    &    &      &        &       &\multicolumn{2}{l}{\it RRE~ 1215}\\
$~~0$ & {\bf 444a}  & 26.1 &  .957 &  .240 &  1.14 & 68.5  & 0.00  & 36.4 &$ ~0$ & {\bf 418a} &  27.7 &  .872 &  .253 &  1.37 & 33.4 & 0.00 & 14.4\\
$-1$&  {\bf 77a} & 150  & .948 &  .333  & 1.41 & 38.7 &  0.00 &  46.6 &$-1$ &{\bf 449a} & 25.8 &  .861 &  .303 &  1.52 & 27.6 &  0.00 & 24.9\\
$-2$ &  {\bf 416a} & 27.8 &  .938 &  .206  & 1.61 & 35.2  & 0.00  & 55.6 &$ -2$ &  {\bf 76a} & 152 &  .845 &  .349 &  1.74 & 24.9 & 0.00 &  34.3\\
$-3$ & {\bf 104a} & 111 &  .935  & .261  & 1.74 & 21.8  & 0.00  & 60.6 &$    -3$ & {\bf  59a} & 196 &  .823 &  .261 &  1.88 & 15.7 &  0.00 &  40.0\\
$     -4$ & {\bf  457a}& 25.3 &  .922 & .159 &  1.93 & 14.3 &  0.02 & 64.1 &          &       &       &        &        &      &      &       &      \\
$     -5$ & {\bf  64b}& 181 &  .912 & .188 &  1.93 & 14.7 &  0.01 & 67.4 &          &       &       &        &        &      &      &       &      \\
$     -6$ & {\bf  129b}& 89.7 &  .900 & .109 &  2.00 & 14.3 &  0.02 & 70.3 &          &       &       &        &        &      &      &       &      \\
$     -7$ & {\bf  480a}& 24.1 &  .893 & .035 &  1.94 & 15.2 &  0.01 & 73.0 &          &       &       &        &        &      &      &       &      \\
$     -8$ & {\bf  563b}& 20.6 &  .888 & .077 &  1.96 & 12.7 &  0.08 & 75.0 &          &       &       &        &        &      &      &       &      \\
$     -9$& {\bf  499b}& 23.2 &  .882 & .118 &  2.02 & 13.2 &  0.05 & 77.0 &          &       &       &        &        &      &      &       &      \\
$    -10$ & {\bf  385b}& 30.1 &  .874 & .158 &  2.04 & 11.9 &  0.18 & 78.7 &          &       &       &        &        &      &      &       &      \\
$    -11$ & {\bf   49b}& 236 &  .869 & .161 &  2.03 & 10.2 &  0.31 & 80.3 &          &       &       &        &        &      &      &       &      \\
$  -12$ & {\bf  361b}& 32.1 &  .858 & .116 &  1.92 & 12.6 &  0.09 & 81.9 &          &       &       &        &        &      &      &       &      \\
$    -13$ & {\bf  148b}& 78.2 &  .850 & .128 &  1.84 & 12.5 &  0.10 & 83.1 &          &       &       &        &        &      &      &       &      \\
\multicolumn{2}{l}{{\it SVC~ 1620}}   &    &     &    &    &      &        &       &\multicolumn{2}{l}{\it RRE~ 1620}\\
$~0$ & {\bf 444a} &  26.1 &  .955 &  .250 &  1.44 & 47.7 & 0.00  & 38.6 &$ ~0$ & {\bf 419a}  & 27.6 &  .831 &  .149  & 1.89 & 21.8 & 0.00 & 12.2\\
$-1$ & {\bf 77a} & 150 &  .946 &  .329  & 1.68 & 33.6 &  0.00 &  48.9 &$-1$ & {\bf 78a} & 148 &  .816 &  .181 &  2.03 & 22.6  & 0.00  & 23.8\\
$-2$ &{\bf 416a} &  27.8 &  .934 &  .196 &  1.93 & 25.0  & 0.00 & 56.3 &$    -2$ & {\bf 450a} & 25.7 &  .787 &  .061 &  2.26 & 16.1  & 0.00  & 31.9\\
$-3$ & {\bf 128a}  & 90.4 &  .930 &  .239  & 2.03 & 17.8 &  0.00 & 61.0 &    &      &     &      &        &      &      &       &     \\
$     -4$ & {\bf  105b}& 110 &  .920 & .064 &  2.17 & 12.3 &  0.12 & 64.2 &          &       &       &        &        &      &      &       &      \\
\multicolumn{2}{l}{{\it SVC~ 2800}}   &    &     &    &    &      &        &       &\multicolumn{2}{l}{\it RRE~ 2800}\\
$~~0$ & {\bf 443a} &  26.1 &  .927 &  .251 &  1.47 & 41.0 & 0.00  & 37.9 &$ ~0$ & {\bf 78a} & 148 &  .775 &  .178 &  1.94 & 21.5 & 0.00 &  11.8\\
$-1$ & {\bf 77a} & 150 &  .917 &  .315 &  1.65 & 34.3  & 0.00 & 47.8 &$ -1$& {\bf 421a} &  27.5 &  .744 &  .066 &  2.13 & 16.7 &  0.00 &  20.7\\
$-2$ & {\bf 416a} &  27.8 &  .900 &  .180  & 1.93 & 19.2 & 0.00  & 53.5 &        &            &      &       &       &       &       &      &     \\
$-3$ & {\bf 128b} & 90.4&  .893 &  .207 &  2.09 & 13.4  & 0.04  & 57.2 &      &      &     &       &       &        &     &       &      \\
$     -4$ & {\bf  430b}& 26.9 &  .883 & .146 &  2.08 & 12.2 &  0.12 & 60.6 &          &       &       &        &        &      &      &       &      \\
$     -5$ & {\bf  487a}& 23.8 &  .876 & .159 &  1.97 & 15.2&  0.01 & 64.1 &          &       &       &        &        &      &      &       &      \\
$     -6$ & {\bf  66b}& 175 &  .868 & .200 &  2.07 & 14.0 &  0.02 & 67.7 &          &       &       &        &        &      &      &       &      \\
$     -7$ & {\bf  52a }& 223 &  .854 & .112 &  2.16 & 15.6 &  0.00 & 71.2 &          &       &       &        &        &      &      &       &      \\
\multicolumn{2}{l}{{\it SVC~ 4995}}   &    &     &    &    &      &        &       &\multicolumn{2}{l}{\it RRE~ 4995}\\
$ ~~0$& {\bf 442a} & 26.2 &  .891 &  .182  & 1.75 & 31.6 &  0.00  & 37.5 &$ ~0$ &  79a & 147 &  .755 &  .009  & 2.66  & 9.1 & 2.78  &  7.1\\
$-1$ & {\bf 77a} & 150 &  .870 &  .221 &  2.00 & 21.3 &  0.00 &  45.3 &     &             &      &       &       &       &      &       &     \\
$-2$ & {\bf 420b} &  27.6 &  .859 &  .114 &  2.20 & 11.3  & 0.33 & 49.4 &       &            &     &       &       &       &       &      &      \\
$-3$ & {\bf 103b} & 112 &  .856 &  .132 &  2.34 & 11.0 &  0.39 & 53.4  &         &       &      &       &       &       &      &       &     \\
\multicolumn{2}{l}{{\it SVC~ 8800}}   &    &     &    &    &      &        &       &\multicolumn{2}{l}{\it RE~ 8800}\\
$~~0$ & {\bf 79a} & 147 &  .756 &  .365  & 1.74 & 42.7 &  0.00 &  26.9 &$ ~0$ & {\bf 81a} & 143 &  .726 &  .338  & 1.89 & 33.5 & 0.00  & 19.6\\
$-1$ & {\bf 66a} & 175 &  .696 &  .189  & 2.20 & 25.9 & 0.00  & 40.5 &$ -1$ &{\bf 67a} & 173 &  .668 &  .185  & 2.28 & 20.9 &  0.00  & 31.5\\
     &            &     &       &       &       &      &       &      &$-2$ & {\bf 51a} & 227 &  .617 &  .073  & 2.55 & 11.3  & 0.32 & 37.8\\

\noalign{\smallskip}
\hline
\end{tabular}
\end{scriptsize}
\end{flushleft}
\end{table*}

Since peaks in a periodogram may arise from aliasing or other phenomena not 
present in Gaussian noise (e.g., spectral leakage arising from the spacing of the 
data and from the finite length of the time series), the $FAP$ values alone, are 
insufficient for establishing whether or not strong peaks in a periodogram are 
indeed real periodicities in the time series. Also, some small peaks present in 
the original periodogram can be real in the case when the normalized factor $k$ would 
be too large in consequence of treating real periods as  noise. We test for 
genuine peaks by recomputing  the periodogram after randomising the data on the 
time grid. This procedure (Delache et al. \cite{del85}; \"{O}zg\"{u}\c{c} \& Ata\c{c} 
\cite{ozg94}) maintains the noise characteristic of the time series 
but destroys all coherent signals, especially those with periods longer 
than the chosen cut interval of the data. In our 
time series, we cut the data with a seven day interval. It preservs 
to some extent the correlation characteristics of the data, so
if a period results from a strong correlation inside the data, the 
number of cases in which it is observed should be rather large. We repeated this 
simulation 10000 times, every time computing the number of cases in which the 
recalculated power values for the periods having peaks in the original spectrum are 
equal to or larger than the peaks power of the real data. The results of these 
calculations are presented in Table 4 and 5 as small letters situated after the 
frequency of the peaks found in the original periodograms. The letter `a' indicates 
that, for 10000 simulations, in less then 10 cases the peak value at the given 
frequency exceeded the corresponding peak power of the real data. Successive 
letters mark the intervals for which the probability (calculated from 10000 simulations) 
of obtaining as high a peak as in the original periodogram by chance are as 
follows: `b': 0.1 - 1\,\%, `c': 1 - 5\,\%, `d': 5 - 15\,\%. 

When more than one periodic signal is present in the data, multiple significant 
peaks appear in the periodogram. Alternatively, a true signal at frequency 
$\omega_{\circ}$ can cause peaks in the periodogram at frequencies 
other than $\omega_{\circ}$ because of the 
finite length of the data and irregularities in the data spacing. A useful procedure 
for determining whether any additional peaks with significant false alarm 
probability are physically real is an iterative peak removal technique 
(Delache \& Scherrer \cite{del83}; Horne \& Baliunas \cite{hor86}).
The highest peak in the original data periodogram provides 
the frequency $\omega_{\circ}$ corresponding to the strongest sinusoidal 
signal present in the data. Using the method of least squares the phase and 
amplitude of this sinusoidal signal are fitted from the original data. 
This allows subtraction of the best fitted sine curve from 
the time series and then recalculation of a new 
periodogram. This procedure is repeated as long as the FAP of the peaks 
connected with subtracted sine curves are smaller than 0.5\,\%, producing the  
main periods in each of the analysed time series. To see the effect of removing 
these peaks from the original data, we present in Fig. 3 the  six consecutive 
periodograms, calculated according to this procedure for the case of the minimum 
 $SVC~810$ time series. We see that the original periodogram is dominated by 
a strong peak at frequency 423 nHz ($27\fd4$ ), so the structure of the periodogram 
near this frequency is difficult to recognise. When the sine curve with this 
period is removed from the data, the new period at $29\fd2$ begins to be visible 
(Fig. 3b). Subtracting the next sine curves from the data we come to Fig. 3f
 which presents the periodogram of the last descendant time series having 
period whose peak value gives FAP smaller than 0.5\,\%.
 Comparing all the periodograms seen in 
this figure, we can conclude that the period values indicated by peak positions 
do not change much during the removed procedure, which confirms the significance 
of the periods found. The results 
of the removal procedure are summarised in Table 2 for the minimum and in 
Table 3 for the rising phase. Both tables show for  the original time series  
( marked by 0), and those with successively removed sine curves 
(denoted by -1,~-2,~-3....) such characteristic parameters as: frequency and period of 
the highest peak, the autocorrelation coefficients, the normalization factor $k$, 
the normalized power equal to the Scargle power divided by $k$, the $FAP$ value 
resulting from the normalized power.

\begin{table*}
\caption{ The characteristic periodicities (lines) observed in the minimum.}
\label{tab4}
\begin{flushleft}
\begin{scriptsize}
\begin{tabular}{cccccccccccccccc}
\hline\noalign{\smallskip}

$ MMF$ &$ SVC$   &$ SVC$  & $SVC$& $ SVC$ &$ SVC$ &$SVC$ & $SVC$ &$ ISN$ & \multicolumn{2}{c}{Line}  &$ M $&$ SVC $&$ISN$& \multicolumn{2}{c}{Time series}  \\
     & $RRE$ &$RRE$ & $RRE$ &$RRE$ & $RRE$ & $RRE$ &$RRE$ &       & [nHz] &  {\bf [day]}&      &$ RRE $& Imp. & $NP$ & {\boldmath $FAP$}\\
  & 405  & 810  & 1215 & 1620 & 2800 & 4995  & 8800  &   & $\Delta f$ &    &   &     &    &   & { [\bf \% ]}\\
\hline
1       &  2     &  3    &    4  &    5  &  6      &  7     &  8      &    9  &  10    &  11  &  12  & 13  & 14  & 15  &  16\\
\hline
{\bf 858a} & 851c &      &      &     &      & 846d & 845d &     & 854.5 & {\bf 13.5} &  a  &  c  &  -  &\multicolumn{2}{c}{$MMF~$/~0}   \\  
          & 851d &        &         &       &          &            &       &       &     13 &           &      & + &  6   &   19.0 & {\bf 0.00}\\
\hline
      & 673c & 679c & 679d & 675c & 674b & 672b &      &      672b  &  673.0 & {\bf 17.2} &  -  &  2b3c& b   &\multicolumn{2}{c}{$ISN~$/-1}  \\
      &       &        &  662c   &  665c   &       &       &  685c &       &  23  &    &   &  3c  &  15  &  6.7 & {\bf 20.93}\\
\hline
         & 588d & 585c & 580b & 584c & 584b  & 585d  &       &  583c  &  586.0  & {\bf 19.8} &    -   & 2b2c &  c  &\multicolumn{2}{l}{$ SVC~ 1215 $/-7}\\
    & 603d &    & 594d  &   & 600c   &   &     &    &  23   &   &    & c  & 10  & 8.4  & {\bf 4.27}\\
\hline
&  534b & {\bf 529b}  & {\bf 528b}  &  526b &  526b  & 525c  &     &  529b &  528.1 & {\bf 21.9} &  -  & 5bc & b  &\multicolumn{2}{l}{$ SVC~ 1215 $/-3}\\
   &  548d &    &   &       &      &       &    &    &  23 &   &    & + & 19  & 12.1 & {\bf 0.11}\\
\hline
         &        &  491c  & {\bf 498b}  &             &         &        &      &          & 496.8  & {\bf 23.3} &  -  &  bc   &   -  &\multicolumn{2}{l}{$ SVC~ 1215 $/-5}\\
       &         & 512d & 503d  & 504c  &   496b  & 495b & 497d  &      &  21  &   &   & 2bc &  11 & 11.8 & {\bf  0.15}\\
\hline
{\bf 419a} & {\bf 418a} &{\bf 423a} &{\bf 424a}  & {\bf 425a} &{\bf 424a} &      424b  &        &      432a  &  423.5  & {\bf 27.3} &  a & 5ab & a  &\multicolumn{2}{l}{$ SVC~~810 $/~0}\\
     &  {\bf 416a} &{\bf 418a} & {\bf 420a} & {\bf 420b} & 434c & 432c &    &   &  18  &   &   & 3ab2c & 58 & 49.9 & {\bf 0.00}\\
\hline
{\bf 400a}  &      &{\bf 396a} &  397b  &  398c &  398c   &      &      &      & 397.8 & {\bf 29.1} & a  & ab2c & - &\multicolumn{2}{l}{$SVC~~810 $/-1}\\
    &   &        &   &   &    &    &        &   &  4   &   &    &  - & 15  & 18.7 & {\bf 0.00}\\
\hline
       & {\bf 390a} & 374c  & 372d   &     &    &   &       &   &  382.6  & {\bf 30.3} &  -  & ac & -  &\multicolumn{2}{l}{$ RRE~~810 $/-2}\\
    &       382a   &{\bf 381a} &{\bf  383b}  & {\bf 382b} &     383b  &  386b &    &   &  18  &  &   & 2a4b  & 28  & 15.3 & {\bf 0.00}\\
\hline
    &   &  307c & 306c &      &  308d &    &       &      &  299.5 & {\bf 39} &  - & 2c & -  &\multicolumn{2}{l}{$ SVC~~810 $/-6}\\
    &   &      &      &  295d &  293c &      292b  &    &   &  16  &   &   & bc  & 6  & 8.2  & {\bf 5.14}\\
\hline
286d  &   &  272c & 265d  & 275c &   &  281d & 265d &   &  270.6 & {\bf 43} &  +  & 2c  & -  &\multicolumn{2}{l}{$ RRE~~810 $/-3}\\
    &   & {\bf 268b} & {\bf 267b} & 266b &    &   &   &   &  21  &   &   & 3b  & 11  & 11.8  &{\bf 0.14}\\
\hline
    &        &  {\bf 238b} & {\bf 242b}& 238d &      &        &    &   & 240.0 & {\bf 48} &  -  & 2b  & -  &\multicolumn{2}{l}{$ SVC~ 1215$/-6}\\
    &   &      &   &   &   &       &   &   &  4 &  &   & -  & 6 & 11.8 & {\bf 0.14}\\
\hline
   &  197d & 195c  &  197c &       &  195c &  198c &  217d  &  194c &  199.0   &  {\bf 58} & -  & 4c & c  &\multicolumn{2}{c}{$ RRE~ 8800$/-3}\\
   &   &       &      & 217d  &    & 204d   & 215c   &   &  23  &    &   & c  & 6  & 8.9 & {\bf  2.55}\\
\hline
     &        & {\bf 150b} &{\bf 151b} &  150c & 147d & 156c & {\bf 154b} &   &  154.4 & {\bf 75} & -  & 3b2c & - &\multicolumn{2}{l}{$ RRE~~810 $/-1}\\
   & 143d & {\bf 152a} &{\bf 157a}  &{\bf 155b} &     159b  &{\bf 158a}  &{\bf 156b}  &   & 16  &   &   & 3a3b & 35  & 27.7  & {\bf 0.00}\\
\hline
   & 125d   &      &    &     &  121b  &{\bf 124a} & {\bf 122b}  &  126c & 126.4 & {\bf 92} & -  & a2b & c  &\multicolumn{2}{l}{$ SVC~ 8800 $/~0}\\
   &   &       & 136c & 135c  &  139d &    &  {\bf 121b} &   &  18 &  &  &  b2c  & 17 & 15.8 & {\bf 0.00}\\
\hline
    &       & 107c & {\bf 113b} &       &        &        &    &  &  112.7  & {\bf 103} &  - & bc & -  &\multicolumn{2}{l}{$ SVC~ 1215 $/-4}\\
    & 118d  & 116d & 118c &   &      &       &      &    &  11  &   &   & c  & 5 & 11.5 & {\bf 0.20}\\
\hline
96c   &    &    &    &    & {\bf 85b}  & {\bf 90a}  &     &{\bf 84a}  &  89.0  & {\bf 130} &  c  & ab &a  &\multicolumn{2}{l}{$ SVC~ 4995 $/~0}\\
    &   &     &   &  88d  &       &96b  &  {\bf 83b} &   &  13 &   &   & 2b & 20  & 18.3 & {\bf 0.00}\\
\hline
    &   75b  &  {\bf 76b} &  {\bf 80b} &   80b &    &  74d  &  80c   &    &  77.4 &  {\bf 150} &  -  & 4bc & - &\multicolumn{2}{l}{$ SVC~ 1215 $/-1}\\
    & 69c  &  82c &   &   &    &  67d  &    &   &  13  &   &  & 2c  & 15  &  14.8 & {\bf 0.01}\\
\hline
60d   &       &  55c   & 56c  &       &  58c   &  47c  &  54c &     &  53.0  & {\bf 218} &  +  & 5c  & -  &\multicolumn{2}{l}{$ RRE~ 4995 $/~0}\\
    &       &  52c  &  52c &  50b &      & {\bf 53a}  &    &   &  13  &   &   & ab2c & 15  & 23.5 &  {\bf 0.00}\\

\noalign{\smallskip}
\hline
\end{tabular}
\end{scriptsize}
\end{flushleft}
The first nine columns give frequencies of the lines found in the indicated type of data. 
The small letter following the frequency denotes the interval to which belongs 
the probability to obtain by chance as high a peak at this frequency as in the original 
periodogram. The intervals are as folows: `a': $<0.1$\,\% `b': 0.1 - 1\,\%, `c': 1 - 5\,\%, 
`d': 5 - 15\,\%. The bold letters indicate frequencies of the
 large peaks having $FAP$ values smaller than 0.5\,\%.

Columns 10 and 11 give the mean frequency and mean period of the line. They are 
calculated only from the periods marked by letters `a', `b', `c'. The number 
under the mean frequency is the difference between the highest and the lowest frequency 
from all the periods found. 

The next three columns (12, 13, 14) describe in a shorter way in what type of data 
(magnetic, radio, spot numbers) the given line is visible. We mark this using 
the letters `a', `b', `c', and two symbols `+' and `-'. The symbol `-' means that 
the given line is not noticed in the respective type of data. 
The letters `a', `b', and `c' indicate that the line is recognised 
 and the periods found have the probability level described by these 
letters, while the symbol `+' indicates that in the given type of data are periods 
having the probability level described by the letter `d'. 
The second row in column 14 gives the importance number which indirectly measure 
how large is support for this line in all the analysed time series (see text). 

The last two columns in the first row, determines the time series in which the highest peak 
for the given line have been observed. The second row gives the normalized power 
of this peak as well as the resulting $FAP$ value. 
\end{table*}

\begin{table*}
\caption{Same as Table 4 but for the rising phase data.}
\label{tab5}
\begin{flushleft}
\begin{scriptsize}
\begin{tabular}{cccccccccccccccc}
\hline\noalign{\smallskip}
$ MMF$ &$ SVC$   &$ SVC$  & $SVC$& $ SVC$ &$ SVC$ &$SVC$ & $SVC$ &$ ISN$ & \multicolumn{2}{c}{Line}  &$ M $&$ SVC $&$ISN$& \multicolumn{2}{c}{Time series}  \\
     & $RRE$ &$RRE$ & $RRE$ &$RRE$ & $RRE$ & $RRE$ &$RRE$ &       & [nHz] &  {\bf [day]}&      &$ RRE $& Imp. & $NP$ & {\boldmath $FAP$}\\
  & 405  & 810  & 1215 & 1620 & 2800 & 4995  & 8800  &   & $\Delta f$ &    &   &     &    &   & { [\bf \% ]}\\
\hline
1       &  2     &  3    &    4  &    5  &  6      &  7     &  8      &    9  &  10    &  11  &  12  & 13  & 14  & 15  &  16\\
\hline
{\bf 876a} &       &      & 887b  & 887d & 886d &       &   & 886c &  883.4  & {\bf 13.1} &  a  & b & c  &\multicolumn{2}{c}{$ MMF~$/-2}\\
      &        & 886d  &  886d  &  887c  &        &  881c  &       &     &  11  &   &  & 2c & 11 & 12.2  & {\bf 0.14}\\
\hline
{\bf 844a} &      & 863c  & 864c   &     &     &     &        &       & 853.2 &  {\bf 13.6} &  a  & 2c & - & \multicolumn{2}{c}{$ MMF~$/-1}\\
      &   &     &     &    &  841c &        &    &    & 23  &    &    &  c  &  8  & 26.9  & {\bf 0.00} \\
\hline
814a &       & 823d  &  822d   &      &        &      &   &     &  806.5*  & {\bf 14.4} &  a  & + & -  & \multicolumn{2}{c}{$ MMF~$/~0}\\
 {\bf 799a}  &       &       &        &        &     &    &  820d &    &  24    &   & a  & +  & 10 & 25.8 & {\bf 0.00}\\
\hline
756a  &        &    &     &     &     &     &        &        &  762.4  &  {\bf 15.2}  &  a  & -  & -  & \multicolumn{2}{c}{$ MMF~$/-3}\\
       &  765c   &        &       &        &  767c   &  763c  & 761c  &   &  11  &    &   & 4c  &  9 & 7.9  &  {\bf 8.86}\\
\hline
    &  565b   &  {\bf 562b}  &  {\bf 563b} &  566c &  571b & 572b &  572d  & 571c  &  569.3 & {\bf 20.3} &  -  & 5bc  & c  & \multicolumn{2}{l}{$ SVC~~810 $/-9}\\
   &  566b  &  572c  &  574b  & 574b & 579d &  576c&  575d  &   &  17  &   &    & 3b2c & 28  & 12.8 &  {\bf 0.07}\\
\hline
      &  502c & {\bf 498b}  & {\bf 499b} &    &  504b &  497a  & 499b & 507d  &  499.3  &  {\bf 23.2} & -  & a4bc & +  & \multicolumn{2}{l}{$ SVC~1215 $/-10}\\
   &  504c &        &       &      & 498d  &  496c  & 498c  &  &  11  &   &   &  3c  & 21  & 13.2 & {\bf 0.05}\\
\hline
   &   &  {\bf 477a} & {\bf 480a}  &  483b &  {\bf 487a}  &    &    & 486b  &  482.6 & {\bf 24.0} &  -  & 3ab  & b  & \multicolumn{2}{l}{$ SVC~1215 $/-7}\\
    &    &    &   &    &        &    &       &   &  10  &   &    & -  & 21  & 15.2  & {\bf 0.01}\\
\hline
   &        & {\bf 456a } & {\bf  457a} &  457b  & 464d  & 453c  &    & 455d &  459.5 & {\bf 25.2} &  - & 2abc & +  & \multicolumn{2}{l}{$ SVC~~810 $/-3}\\
    &        & 467b  &  466d & 467c &     &       &        &   &  14  &   &    & bc  & 18 & 20.1  & {\bf 0.00}\\
\hline
444d  &  {\bf 444a}  & {\bf 443a} & {\bf 444a} & {\bf 444a}  &{\bf 443a} & {\bf 442a} &  439b  & {\bf 442a}  &  444.4  &  {\bf 26.0} & +  & 6ab & a &\multicolumn{2}{l}{$ SVC~~810 $/~0}\\
   &        & {\bf 450a} &{\bf 449a}  &{\bf 450a}  &         & 443b &    &   &  11 &   &   & 3ab  & 56 & 74.0  &  {\bf 0.00}\\
\hline
    &        & 429b  & 429b & 432b &  {\bf 430b} &  431d &    &   &  431.7 & {\bf 26.8}  &  -  & 4b  & -  &\multicolumn{2}{l}{$ SVC~2800 $/-4}\\
   &   &  434d & 434c & 436d &   436b  &    &      &    &  7   &   &   &  bc  & 16  & 12.2  & {\bf 0.12}\\
\hline
   & {\bf 418a}  &{\bf 416a} & {\bf 416a}  & {\bf 416a}  & {\bf 416a}  &{\bf 420b} &  415c  &  414c &  417.4 & {\bf 27.7} &  - & 5abc & c &\multicolumn{2}{l}{$ RRE~~810 $/-1}\\
   &  {\bf 418a} &{\bf 417a}  &{\bf 418a}  &{\bf 419a}  &  {\bf 421a}  &  420a &        &   &  7  &    &    & 6a  & 60  & 50.4  & {\bf 0.00}\\
\hline
399a   &      &   &   &     &     &      &    &    &  399.0 & {\bf 29.0} &  a  & -  & - &\multicolumn{2}{c}{$ MMF~$/-3}\\
    &   &    &  399d  & 400c & 398c &      &    &  &   2 &    &    & 2c & 7 & 10.8  &  {\bf 0.54}\\
\hline
377a   &  382c   &  {\bf 385a} &  {\bf 385b} & 383c  &        &       &  376d & 382d & 381.1  & {\bf 30.4} &  a  & ab2c &  + &\multicolumn{2}{l}{$ SVC~~810$/-7}\\
    &  387d & 381b  &  381c & 381c  &  383c & 383d &  373c  &   & 14  &    &    & b4c & 22  & 15.4 & {\bf 0.01}\\
\hline
     &     & 362b  &  {361b}  &  366d &  366c &        &       &      & 363.8  & {\bf 31.8} & -  & 2bc  & -  &\multicolumn{2}{c}{$ SVC~ 1215$/-12}\\
     &   &       &   & 365d  &  363c  &      &      &    &   5  &   &    &  c  & 8  & 12.6& {\bf 0.09}\\
\hline
     &     & 285d  &  284c  &  284d &  284c &  283d  & 292c  & 285c & 287.0  & {\bf 40} & -  & 3c & c  &\multicolumn{2}{c}{$ SVC~ 1215$/-14}\\
   &   &       &   &   &  281d  &  284d  &  290b &    &  11  &   &    & b   & 7  & 10.3& {\bf 0.85}\\
\hline
157d &  149d   & 146c  &  {\bf 148a}  &  151c &  152d &        &       &      & 150.0  & {\bf 77} & +  & b2c & b  &\multicolumn{2}{c}{$ SVC~ 1215$/-13}\\
   &   &       &   & 141d  &        &         & 155c  &    &  16  &   &    &  c  & 6  & 12.5& {\bf 0.10}\\
\hline
      & 128b  & {\bf 127b} & {\bf 129b}  & {\bf 128a}  & {\bf 128b}  &  127b & 117d  & {\bf 127a} & 126.8 & {\bf 91} &  - & a5b  & a  &\multicolumn{2}{l}{$ SVC~1620 $/-3}\\
   &       &       &       &   &    &    &  120c &   & 12 &   &   & c & 26  & 17.8 & {\bf 0.00}\\
\hline
      &   &  {\bf 103a} & {\bf 104a}  & {\bf 105b}  &  102c & {\bf 103b} &   &     & 103.6  & {\bf 112} & - & 2a2bc & -   &\multicolumn{2}{l}{$ SVC~1215 $/-3}\\
   &   &        & 104c & 104c  & 102d &       &  &    &  3  &  &   &  2c  & 19  & 21.8  & {\bf 0.00}\\
\hline
    &  85a  & 98c &  88d  &       &      &  88d  &  93c  &   &  91.6 & {\bf 126} &  - & a2c & - &\multicolumn{2}{l}{$ SVC~~405 $/-3}\\
    &  86c   &    &   &    &    &    & 96b   &   & 13  &   &    &  bc  & 11  & 9.6 & {\bf 1.70}\\
\hline
       &      & {\bf 75a }  & {\bf 77a} & {\bf 77a}  & {\bf  77a}& {\bf 77a}  &{\bf  79a} & 74b  &  76.8 & {\bf 151}  &  -  & 6a  & b  & \multicolumn{2}{l}{$ SVC~8800 $/~0}\\
      &      &  {\bf 71b}  &  {\bf 76a} & {\bf 78a}  & {\bf 78a}  &  79a  & {\bf 81a}  &      &  10   &   &    & 5ab  &  61 & 42.7 & {\bf 0.00}\\
\hline
       &      & {\bf 62a} &  {\bf 64b}  &  65c  & {\bf 66b}   &  67d  & {\bf 66a}  &     &  64.1  & {\bf 181} &  -  & 2a2bc  & -  &\multicolumn{2}{l}{$ SVC~8800 $/-1}\\
      &     &           &  {\bf  59a}&    65d  &  64b     &     & {\bf 67a}  &     &  8    &   &    &  2ab  & 30  & 25.9  & {\bf 0.00}\\
\hline
     &  {\bf 55a} & {\bf 56a}&  {\bf 49b}  &    &  {\bf 52a} &    & 51b   &    & 52.3   & {\bf 221} &  - & 3a2b  & -  &\multicolumn{2}{l}{$ RRE~~810 $/-2}\\
   &  {\bf 52a}  &         &       &      &      &  54d  & {\bf 51a} &    &   7  &    &   & 2a    & 31  & 20.3 & {\bf 0.00}\\
\noalign{\smallskip}
\hline
\end{tabular}
\end{scriptsize}
\end{flushleft}
*This is the mean value from two close magnetic lines.
\end{table*}

\begin{figure*}
%\vspace{6cm}
\resizebox{\hsize}{!}{\includegraphics{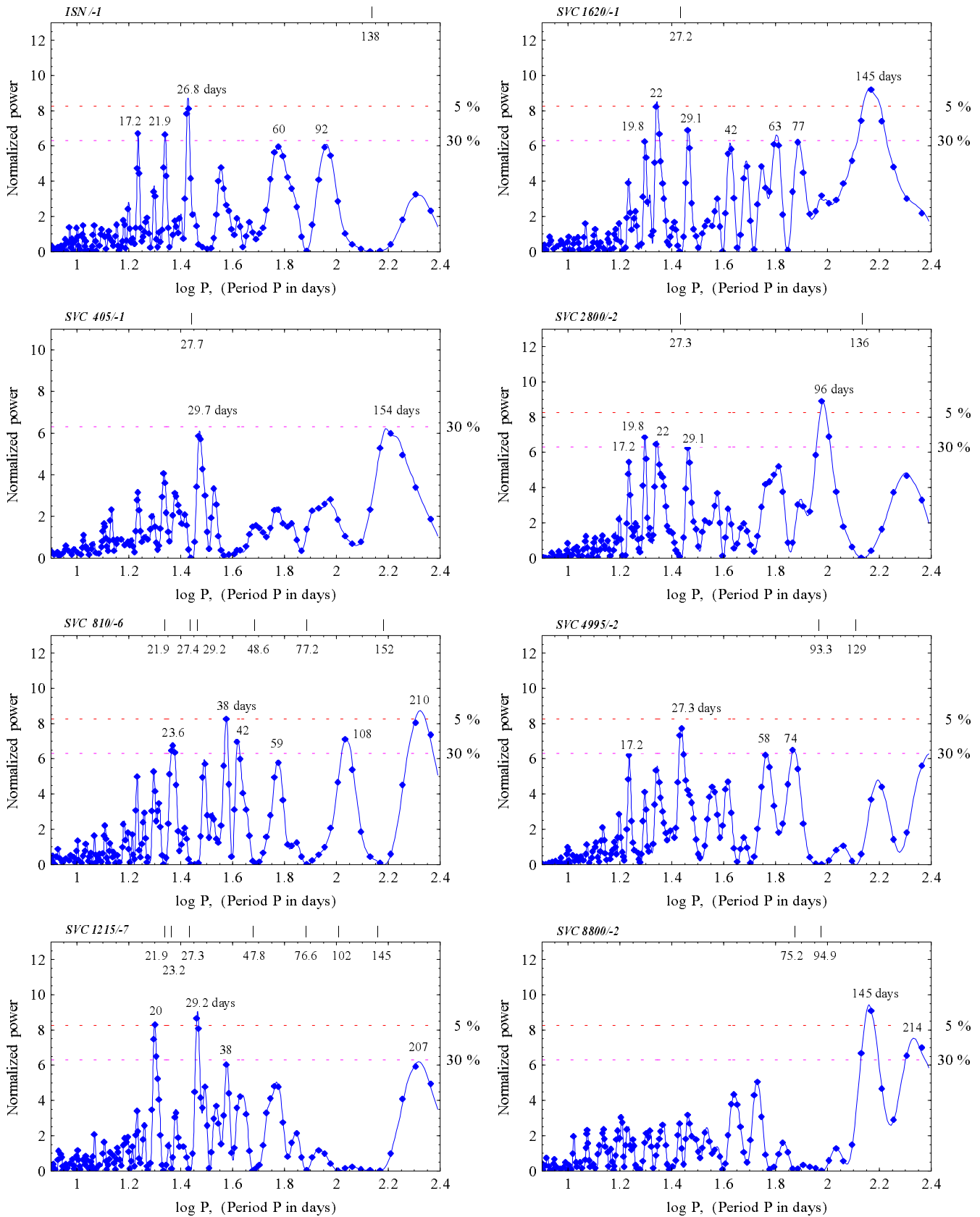}}
\caption{
 The normalised periodograms of the minimum $ISN$ and $SVC$ time series after removing from the original data all the 
 sine curves having periods with $FAP$ values smaller than 0.5\,\%. The removed periods are shown at the top of each graph. The dashed lines indicate $FAP$ 
 significance levels.
}
\label{rw3}
\end{figure*}

\begin{figure*}
%\vspace{6cm}
\resizebox{\hsize}{!}{\includegraphics{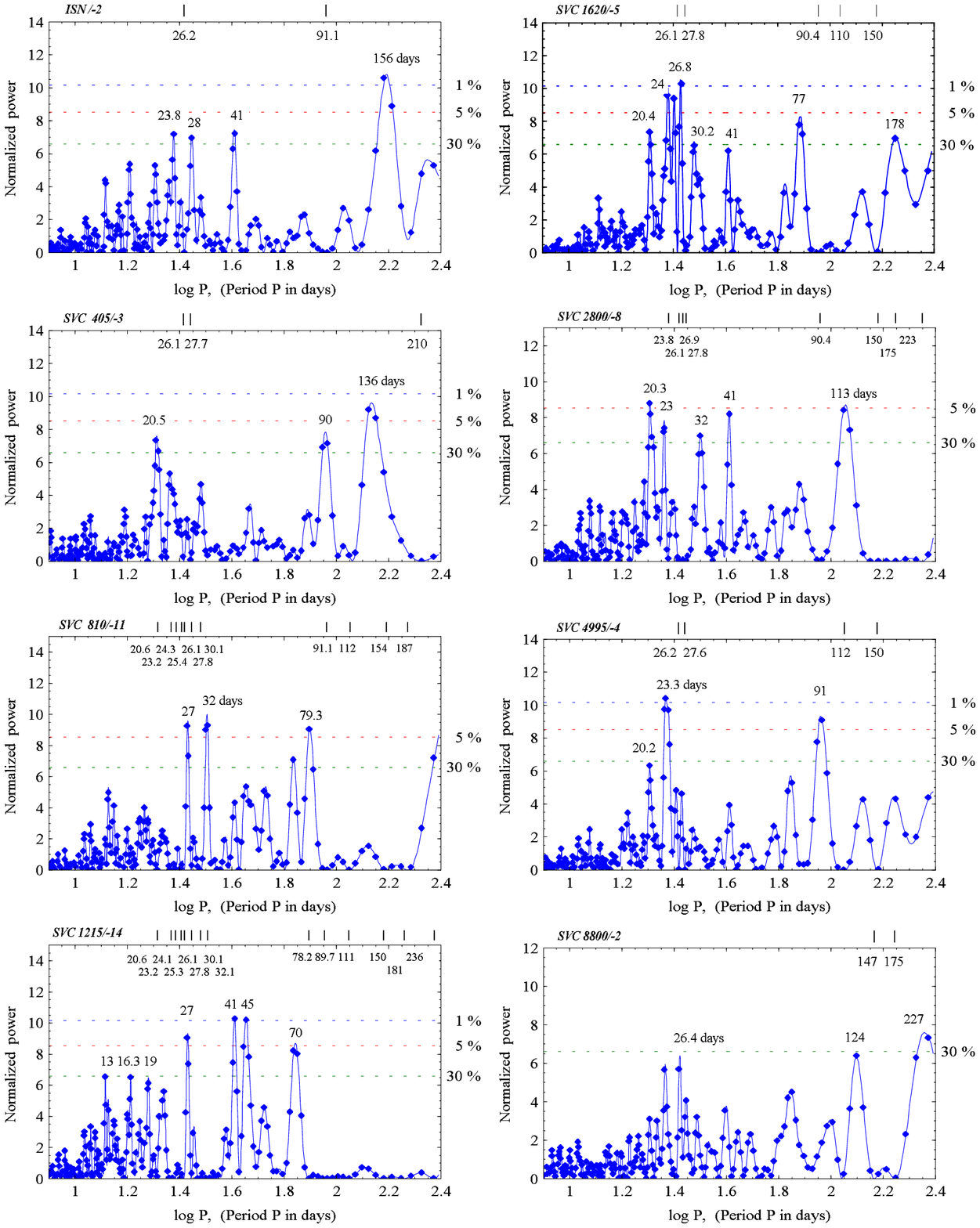}}
\caption{
Same as Fig. 4. but for the rising phase $ISN$ and $SVC$ time series after 
removing  from the original data all the
 sine curves having periods with $FAP$ values smaller than 0.5\,\%.
}
\label{4w4}
\end{figure*}

\begin{figure*}
%\vspace{6cm}
\resizebox{\hsize}{!}{\includegraphics{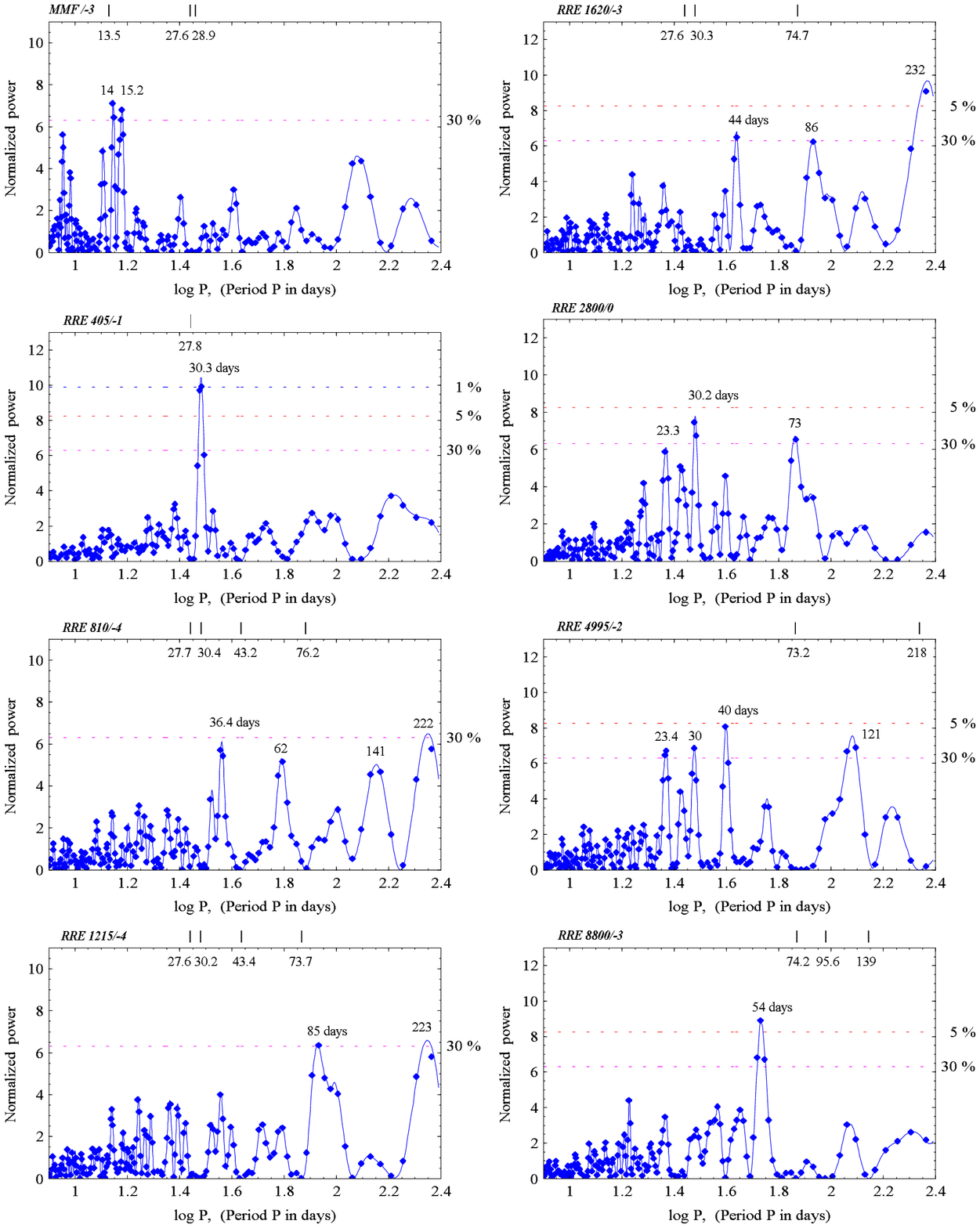}}
\caption{
Same as Fig 4. but for the minimum $MMF$ and $RRE$ time series after
removing  from the original data all the
 sine curves having periods with $FAP$ values smaller than 0.5\,\%.
}
\label{4r3}
\end{figure*}

\begin{figure*}
%\vspace{6cm}
\resizebox{\hsize}{!}{\includegraphics{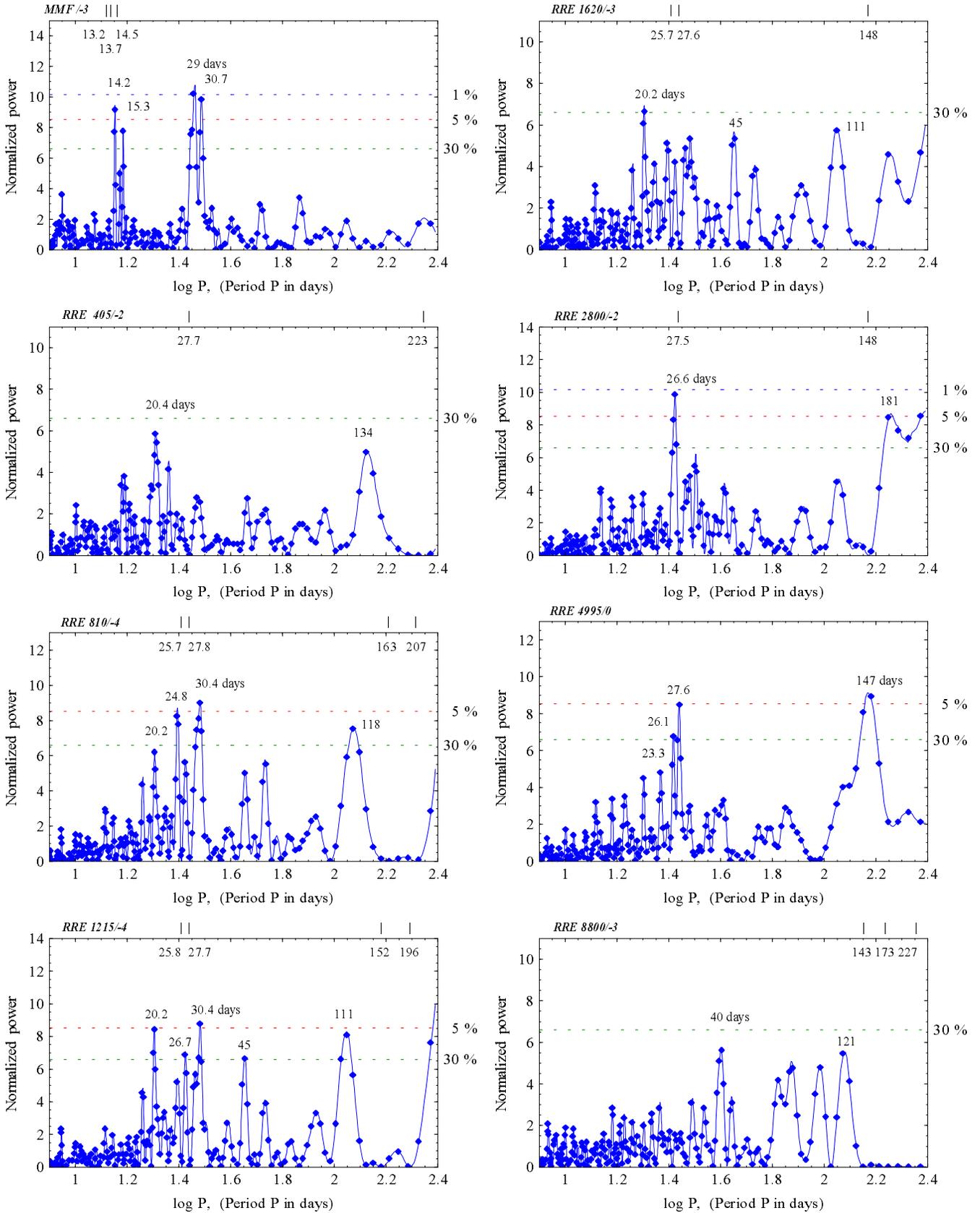}}
\caption{
Same as Fig. 4. but for the rising phase $MMF$ and $RRE$ time series after
removing  from the original data all the
 sine curves having periods with $FAP$ values smaller than 0.5\,\%.
}
\label{4r4}
\end{figure*}

\section{Results and Discussion}

Most of the calculated periodograms have been obtained from radio observations 
($SVC$ and $RRE$ time series calculated for seven different frequencies), but the 
$ISN$ and $MMF$ data were also analysed. The final periodograms obtained 
when all the main sinusoidal signals given in Tables 2 and 3 are subtracted 
from the original data are presented in Fig. 4, 5, 6 and 7. Careful examination of these 
periodograms shows that many periods having a formal significance level near 
30\,\% are seen in time series coming from different observational data. This 
fact allow us to suggest that even those periods could be real. 

Taking into account all the information about the periods found obtained with 
the three different approaches to the data described in Sect.\,3, we prepare 
Table 4 for the minimum, and Table 5 for the rising phase, which bring together 
information about characteristic periodicities (lines) observed in these phases of the solar 
cycle 23 (we use the spectroscopic term `line' for the mean period calculated from 
the periods observed in various time series, which, as we suppose, represent the 
same characteristic periodicity). 
As two time series were created from the radio data at each frequency, 
we use two successive rows in the radio data columns to separate periods 
observed in $SVC$ and $RRE$ radio time series. The exact frequencies of 
the detected periods were found with a 1 nHz resolution in the vicinity 
of peaks seen in periodograms constructed from normalized power values 
on the grid of the independent frequencies appropriate to the given window. 
In all the figures, the points marked in the presented 
periodograms are situated in the places resulting from the grid of the independent 
frequencies, while the solid lines are drawn from the power values computed 
every 1 nHz. 
To discribe in some measure a `strength' of the line we introduced so called 
the importance number of the line. This number shows how large is support 
for this line from all the analysed time series. It is calculated 
according to the following rule. The each letter `a' in the line description 
columns (columns 12, 13, 14) gives for the importance number `5', the 
letter `b' gives `3' , and the letter `c' only `1' . 
It is only one line with the importance number `5' in Tables 4 and 5. There 
are a few lines with a smaller importance number than `5' recognised in the 
analysed data, but we do not include them in Table 4 and 5. 
However, it is important to notice that some of the lines included in Tables 4 and 5
can be unreal in a sense that they are created from periods which in fact belong 
to two different but neighboring lines. For such a case the calculated mean period 
is somewhere between the periods of these two neighboring, unknown lines. 
The probability of such a situation increses for lines having long periods and large
values of $\Delta f$ (Column 10, second row).

To aid in further discussion of the lines we have prepared 
Fig. 8, which shows all the lines present in Tables 4 and 5. 
The level of darkness in this 3D graph illustrate the importance number of the lines.
A close look at Fig. 8 shows a clear difference among the 18 lines observed 
in the minimum and the 22 lines found in the rising phase of solar sunspot activity 
cycle. Although the 9 lines have almost the same periods in both the phases, 
the strength (measured indirectly by the importance number) of nearly all 
the 18 lines change, indicating that perhaps the physical mechanisms responsible 
for them also change with the solar phase. From Fig. 8 it is evident that the lines 
in the rising phase gather into three groups: 
\begin{enumerate}
\item \emph {Magnetic lines:\/} all lines with periods shorter than 16 days. They have small 
importance numbers and were included on our list only because of very high peaks 
observed at these short periods in the $MMF$ (mean magnetic field) time series. In the 
minimum, only one line (13\fd5) from this group is enough strong to be present in 
Fig. 8, but in the minimum $MMF$ periodogram (see Fig. 6) some of them are 
easily seen. 
\item \emph {Rotational lines:\/} all lines with periods inside the 25 - 34 days time 
interval of the Sun's rotation. The three lines from this group are 
visible in the minimum. The line 27\fd3 has the largest importance number and 
results probably from the rotation of such solar phenomena as new solar cycle 
sunspots as well as the long-lived coronal streamer structure observed during the
1996 minimum (Lewis et al. \cite{lew99}). In the rising phase the period of 
this line shifts to 27\fd7. The next two lines 29\fd1 and 30\fd3 are stronger 
in the minimum than in the rising phase. This additionally supports the 
supposition that they are associated with some medium, and large scale magnetic structures 
(coronal neutral sheet, global neutral lines) which dominate in the 
minimum and then decline (Lantos et al. \cite{lan92}; Lantos \cite{lan99}).
In the rising phase four new lines appear (25\fd2, 26\fd0, 26\fd8, 31\fd8). Three have periods 
shorter than those discussed above. One of them, having a period of 26\fd0, 
is the third dominant line of this phase. We suppose that this line is strongly 
connected with eruption of new active-region magnetic fields emerging within
the complexes of activity `active longitudes' (Bumba \& Howard \cite{bum69}; 
Ruzmaikin \cite{ruz98}) and its period may be identified with a fundamental 
period of unknow Sun's clock which value is freqeuntly taken as equal 
to 25\fd5 or 25\fd8 (Bai \& Sturrock \cite{bai93}; Bai \cite{bai94}).   
\item Activity lines: all lines with periods longer than 75 days. We propose this 
name as the strongest 151 days line in this group is thought to be related to the 
well known periodicity near l54 days seen mainly in the occurrence of high energy 
solar flares at  the maximum of the solar sunspot activity. It is interesting 
to notice that this line is also observed in the minimum, but with a small 
importance number and only in the radio time series.
\end{enumerate}
In the minimum, the distribution of the line periods in the investigated time 
window is much more uniform. There are 5 lines with periods shorter than 25 days but 
only one belongs to magnetic lines. Within an accuracy of $\pm0\fd3$ all of 
them were noted before in various time series (Hughes \&Kesteven \cite{hug81}; 
Pap et al. \cite{pap90}).
The lines with mean periods $13\fd6$ and $23\fd2$ are the most interesting. They belong 
to the group of 9 lines which are observed in both phases of the solar cycle 
investigated. Although in our data their importance numbers are not large, 
they were recognised as the dominant periods in the SMM/ACRIM 
total irradiance data for the years 1980 - 1988 (see Pap et al. \cite{pap90}). 
A very intriguing group of lines is formed by 4 lines (39, 43, 48, 58 days) 
also observed in the minimum. All these lines are seen only in the time series 
which were obtained from the radio data, except for one with period 58 days 
also observed in the $ISN$ data. Among them only the line at 39 days is also present 
in the rising phase, although a small shift of its period is visible. 
Coming to longer periods we see that most of them are present in both investigated 
phase of the solar cycle. However, in the rising phase we observed more lines 
and all of them are stronger than those in the minimum.
In the minimum, the strongest line in this group has period 75 days. However, 
in the rising phase this line is more weaker.  It is not clear if it
 should be correlated with periods 
73-78 days observed by some authors (Bai \& Sturrock \cite{bai91}; 
Bai \cite{bai92a}; \"{O}zg\"{u}\c{c} \& Ata\c{c} \cite{ozg94}) mainly in flare 
activity during the maximum of a solar cycle. We suppose that this line could be 
important if it is a real harmonic of the best known line with period near 154 days. 
 It is important to notice 
that all of them were mentioned before in various analyses 
(Lean \& Brueckner \cite{lea89}; Pap et al. \cite{pap90}; Bai \& Sturrock 
\cite{bai91}; Kile \& Cliver \cite{kil91}; Antalov\'{a} \cite{ant99}). 

Many previous studies by a number of authors have resulted in a wide range of solar 
periodicities, which are not easy to explain. This indicates that the problem of 
solar periodicities is still open and more systematic efforts should be undertaken. 
Here, we do not want to discuss all possible causes of the observed periods, but we 
want to present a suggestion which may be of help in further investigations. 

Recently Oliver et al. (\cite{oli98}) proposed that the periodic emergence of magnetic flux, 
manifested as sunspots, triggers the near 158 day periodicity in high-energy solar flares. 
As different magnetic features have different rates of rotation (Gilman \cite{gil74}; 
van Tend \& Zwaan \cite{van76}; Erofeev \cite{ero99}) 
we think that a periodic emergence and a constant conversion of various magnetic 
structures explain the origin of the observed lines and their transformation with 
the phase of solar cycle 23.
The main arguments supporting this idea are as follows: 
\begin{enumerate}
\item There is a clear difference between the lines obtained in the minimum and in 
the rising phase. The lines characterictic for the minimum are probaly connected with 
both the short-lived small-scale magnetic fields which originate fairly high in the 
convective zone (Golub et al. \cite{gol81}; Rivin \cite{riv99}) and large-scale fields 
structures like coronal holes, coronal neutral sheet, global neutral lines. All these 
small and large scale magnetic fields dominate in the minimum phase of solar cycle 
(Golub et al. \cite{gol81}; Wang et al. \cite{wan96}; Lantos \cite{lan99}).
\item In the rising phase a new distinct rotational line at 26.0 days is seen. This period 
is equal to the rotation rate of the active longitudes (zones) at 30\degr (where the new flux 
appeared) determined from SOHO/MDI magnetograms by Benevolenskaya et al. 
(\cite{ben99}). 
They found in the northeren hemisphere the rotation rate 446.6 $\pm$ 1.7 nHz and 
444.8 $\pm$ 1.6 nHz in the southern one, while our value for the whole Sun data is 
444.4 $\pm$ 4 nHz. Also the rotation rate of the equatorial zons equal 461.8 nHz from 
SOHO observations is almost the same as  the period 25\fd2  (459.5 nHz) seen in 
our radio data (see Table 5).
\item The line at 151 days is in the rising  phase very prominent. This period is 
connected by Bogart (\cite{bog82}), Bai (\cite{bai87}) and 
Oliver et al. (\cite{oli98}) with strong, long-lived active regions giving 
most of the energetic solar events. This regions  are generated by the 
emergence within the `active longitudes' magnetic tubes formed in the 
dynamo located at the base of the convection zone (Golub et al. \cite{gol81}; 
Benevolenskaya et al. \cite{ben99}; Rivin \cite{riv99}).
We also observe in the rising phase $ISN$ periodogram periods such as 
13\fd1or 23\fd8, which were  linked with the `active longitudes' and active 
centers in deep layers by Bai (\cite{bai87}).
    \end{enumerate}
	
\section{Conclusion}

\begin{figure}[t]
%\vspace{6cm}
\resizebox{\hsize}{!}{\includegraphics{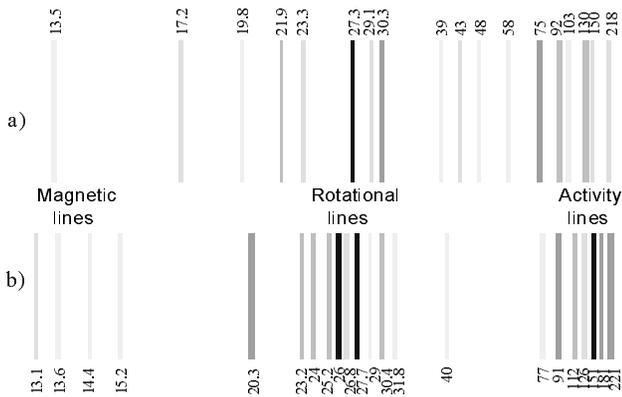}}
\caption{
Characteristic lines generated by solar phenomena in  a) the minimum and  b) the rising phase 
of solar cycle 23. The three group of lines are inicated: The level of 
darkness qualitatively illustrate the importance of lines.
}
\label{lines}
\end{figure}

To summarise, our analysis of radio observations gives the majority of the 
known periods, reveals a clear difference among periodicities observed in the two
phase of the solar cycle examined and shows that daily measured radio fluxes at 
various frequencies are very useful for the systematic study of solar periodicities 
observed in the different layers of the solar atmosphere. We are preparing a 
similar analysis for the next phases of solar cycle 23: the maximum and the 
declining phase. 

We found the rotation rate of the `active longitudes' in the rising phase as equal 
to 444.4 $\pm$ 4 nHz (26\fd0 $\pm$ 0\fd3). We suppose that this period can
be identified with a fundamental period of unknow Sun's clock as a lot of the known 
periodicities are subharmonics of it. We think that in the minimum
the lines observed are conected with the small and large-scale 
magnetic fields which then dominate , while in the rising 
phase most lines are generated by new magnetic structures connected with long 
lived active regions formed within `active longitudes'.

To understand the cause of all the observed periodicities, we still 
need new observational data, but more careful analyse of old radio data can also 
be useful (in preparation). However, our investigation, idicates that the solution 
of the observed solar periodicities should be sought in a complicated Sun's 
magnetic system which generate in the different solar data the compound set of 
solar periodicities.
	
  \begin{acknowledgement}
First of all, we would like to thank our anonymous referee for very valuable 
remarks and suggestions which helped us to significantly improve the current 
version of the paper. The autors are grateful to Dr K. Chy\.{z}y for assistance 
in preparation of the manuscript.
This work was supported in partly by KBN grant No.158/E - 338/SPUB - 204/93.
  \end{acknowledgement}

\end{document}